# Measuring $q_0$ from the Distortion of Voids in Redshift Space

Barbara S. Ryden [1]

Department of Astronomy, The Ohio State University,

174 W. 18th Ave., Columbus, OH 43210

## ABSTRACT

Because the transformation from distance to redshift is nonlinear, maps in redshift space become increasingly distorted as the redshift $z$ becomes greater. As noted by Alcock & Paczyński (1979), observed redshift distortions can be used to estimate the deceleration parameter $q_0$. If $q_0$ is greater than $-1$, voids in redshift space will be elongated along the line of sight. In addition, distant voids will have a greater volume in redshift space than nearby voids. Accurate measurement of the volume and the axis ratio of voids, as a function of their central redshift, will provide an estimate of $q_0$.

To test this method of estimating $q_0$, I create a two-dimensional toy universe, free of peculiar velocities, in which the galaxies are located near the walls of Voronoi cells. The galaxies are then mapped into redshift space, adopting different values of $q_0$. In redshift space, I estimate the area and the axis ratio of the voids by fitting ellipses within the voids, using an algorithm which maximizes the area of the empty ellipses and ensures that ellipses do not overlap. The accuracy of the estimated values of $q_0$ is limited by the intrinsic scatter in the size and shape of the voids. In the toy universe, distinguishing between a $q_0 = -1$ universe and a $q_0 = 1/2$ universe requires a survey which goes to a depth $z \gtrsim 0.1$ in redshift space. Peculiar velocities will create an additional source of uncertainty for the values of $q_0$ measured in N-body simulations and in the real universe.

*Subject headings:* cosmology: theory – galaxies: distances and redshifts – large-scale structure of universe

---

[1]National Science Foundation Young Investigator; ryden@mps.ohio-state.edu

## 1. Introduction

A quarter-century ago, cosmology was characterized as the search for two numbers: $H_0$ and $q_0$ (Sandage 1970). Although the scope of cosmology has broadened, those same two numbers are still being assiduously searched for. I don't intend, at the moment, to plunge into the $H_0$ fray. However, in this paper, I will describe how $q_0$ may be determined (in principle) by measuring the size and shape of voids in redshift space.

The classical tests for $q_0$, as described by Sandage (1988), involve measuring the flux of standard candles or the angular size of standard rulers. The standard candles used to test for $q_0$ are usually galaxies. Accurately determining $q_0$, therefore, requires knowing how the intrinsic luminosity and color of galaxies evolve with time. Unfortunately for cosmologists, galaxy evolution is a complex process. Uncertainties in the evolutionary models lead to a large uncertainty in the derived values of $q_0$. Among the objects used as standard rulers to determine $q_0$ are brightest cluster galaxies (Sandage 1972), extended radio sources (Miley 1971, Kapahi 1989, Daly 1994), and clusters of galaxies (Hickson & Adams 1979, Bruzual & Spinrad 1978). Once again, the difficulty of determining the intrinsic evolution results in a large uncertainty in the value of $q_0$ which is determined.

It is desirable, therefore to measure cosmological parameters by using method which are uncontaminated by evolutionary effects. One such method was proposed by Alcock & Paczyński (1979), who proposed measuring the width and depth, in redshift space, of objects which are intrinsically spherical (or nearly so) in real space. The distortion of the spheres in redshift space would provide an estimate of $q_0$ and the cosmological constant $\Lambda$. At the time of publication, Alcock and Paczyński had no data to which they could apply their test, so their paper remained an undercited curiosity. Recently, however, interest in their geometrical method has revived. Phillips (1994) has proposed using quasar pairs (assumed to be randomly oriented in space) as a substitute for the intrinsically spherical objects of Alcock and Paczyński. Placing interesting limits on $q_0$ using quasar pairs will require observations of approximately 1000 pairs at redshifts $z > 1$. In this paper, I will apply the method of Alcock & Paczyński to galaxies at relatively small redshift ($z \lesssim 0.25$). I will describe how the size, as well as the shape, of voids in redshift maps can be used to estimate $q_0$. In section 2, I describe how measurements in redshift space can be used, in principle, to measure $q_0$. In section 3, I explain a practical method for measuring the shapes of voids, and apply it to simple "toy" universes. Section 4 contains a discussion of the practicality of applying the void distortion test to real redshift surveys.

## 2. Theoretical Expectations

Consider the idealized case of a spherical void which is surrounded by a thin wall of galaxies. At a time $t$, the void wall has a radius $r_v(t)$, and the galaxies within the wall are moving outward with a velocity $u_v(t)$. At some later time $t_0$, an observer located outside the void collects light that was emitted from the void walls at the time $t$ (plus or minus the light travel time across the void). The redshift of a galaxy on the near side of the void is $z - \delta z$ and the redshift of a galaxy on the far side is $z + \delta z$, where $z$ is the redshift of the void center and

$$\delta z = (1+z)\frac{u_v(t)}{c} \ . \tag{2.1}$$

The angular radius of the void as seen on the sky is

$$\delta\theta = \frac{H_0 r_v(t)}{c}\frac{1+z}{y(z)} \ , \tag{2.2}$$

where $H_0$ is the Hubble constant at time $t_0$ and $y(z)$ is the (dimensionless) angular size distance (Peebles 1993). In a spatially flat universe, the angular size distance is

$$y(z) = H_0 \int_0^z \frac{dz}{H(z)} \ . \tag{2.3}$$

If the void is plotted in redshift space, the ratio of its depth to its width is

$$e_v(z) \equiv \frac{\delta z}{z\delta\theta} = \frac{u_v}{H_0 r_v}\frac{y(z)}{z} \ . \tag{2.4}$$

In general, the function $e_v$ is not equal to one, even for a void which is intrinsically spherical and which is simply expanding along with the Hubble flow.

If the galaxies in the void wall are expanding along with the Hubble flow, then $u_v = H r_v$. However, voids tend to expand more rapidly than the Hubble flow. Thus, for a spherically expanding void,

$$u_v(t) = H(t) r_v(t) \left[1 + F(t)\right] \ , \tag{2.5}$$

with the excess expansion factor $F$ being greater than zero for voids expanding more rapidly than the Hubble expansion. In a spatially flat, matter-dominated universe, an isolated spherical void of finite radius evolves into a self-similar state with $F = 1/5$ if the void is compensated (that is, if the net mass deficit is equal to zero), and $F = 1/3$ if the void is uncompensated (Bertschinger 1985). If the initial underdensity $\delta$ of the void has a power-law form, $\delta \propto -r^{-\nu}$, then $F = 1/\nu$ when $2 \leq \nu \leq 3$ (Fillmore & Goldreich 1984). When the initial density profile has $\nu < 2$, then the void wall is a density wave, and matter moves outward with $F \approx 1/2$ (Ryden 1994). If the universe is not flat, then $F$ is a function of the density parameter $\Omega$. In an open universe, Regös and Geller (1991) find that compensated voids have $F \approx \Omega^{0.6}/5$.

Thus, an isolated void, when plotted in redshift space, will be stretched in the radial direction as a result of its peculiar expansion velocity, in addition to the distortion which results from cosmological factors. The observed distortion of a void will be

$$e_v(t) = \left(\frac{H(t)}{H_0}\frac{y(z)}{z}\right)[1 + F(t)] , \qquad (2.6)$$

with the term in parentheses representing the cosmological distortion and the term in square brackets representing the distortion due to the peculiar velocity of the void's expansion.

In order to determine cosmological parameters from the observed distortion of voids in redshift space, I must somehow distinguish between the cosmological distortion and the peculiar distortion. Observed voids are not isolated structures. The distribution of galaxies in redshift space is 'bubbly'; galaxies are typically located in relatively narrow walls between voids which are nearly space-filling. The average void size, when measured in comoving units, can only increase by the merger of two adjoining voids to form a single void, or by the expansion of one void at the expense of a neighboring void. Consider two adjacent voids, with radii $r_1$ and $r_2$. If $r_1 = r_2$, then the wall between the voids will have no peculiar velocity in the direction perpendicular to its surface; if $r_1 > r_2$, the void wall will be pushed in the direction of the smaller void, with an excess velocity (Regös & Geller 1991)

$$F \approx (1 - r_2/r_1)\Omega^{0.6}/5, \qquad (2.7)$$

compared to the Hubble expansion velocity $Hr_1$. In addition to the velocity perpendicular to the void wall, the galaxies within the wall tend to flow outward within the plane of the wall. The pair of neighboring voids is thus converted into a single void, as the intervening wall breaks up. The velocities tangential to the void wall have a maximum excess velocity $F \sim 0.1 - -0.3$, when compared to the Hubble expansion velocity $Hr_1$ (Dubinski et al. 1993).

In a hierarchical scenario, the universe, at any given time, is filled with voids of a characteristic size $R_v$ (Dubinski et al. 1993). Since all voids are roughly the same size, the peculiar velocities in the direction perpendicular to void walls, as found from equation (2.7), should be small, even if $\Omega$ is close to one. The peculiar velocities in the direction parallel to the void walls may have values of $F$ as large as 30 percent. However, if the evolution of the hierarchical universe is approximately self-similar, then the average value of $F$ should remain constant with redshift. The average distortion of voids in redshift space due to peculiar velocities will be constant with redshift; the distortion of voids due to cosmological effects will increase with increasing redshift.

Observationally, there is no tendency for nearby voids to show distortions resulting from peculiar velocities. The analysis by Slezak, de Lapparent, and Bijaoui (1993) of the first CfA redshift slice picks out 15 voids whose centers are at a redshift $z \lesssim 0.04$. This small sample of nearby voids shows no tendency to be elongated along the line of sight in a redshift plot, indicating that the peculiar velocities of the void walls cannot be large. The largest void in the slice is located at $z \approx 0.022$, with $\delta z \approx 0.008$. Comparison of the redshifts and Tully-Fisher distances for the galaxies within the walls of this void shows no detectable outflow; the $1\sigma$ upper limit on the peculiar void expansion is $F < 0.05$ (Bothun et al. 1992). However, the sample of voids within the CfA slice is small and is at low redshift. For accurate measurement of the cosmological distortion, a deeper survey is required.

## 3. Measuring Distortions

For simplicity, I am going to measure the distortion of voids in a universe in which the galaxies have no peculiar velocities. Thus, I am assuming that the void walls are fixed in comoving coordinates, and am ignoring the presence of virialized clusters embedded within the void walls. Such clusters, if located on the near or far side of a void, will alter the void shape by sticking 'fingers of god' into the void. A future study (Melott & Ryden 1995) will use n-body simulations to examine the effects of virialization on the shapes of voids in redshift space. In this paper, for the sake of simplicity, I will look at a synthetic universe which is expanding smoothly with the Hubble flow.

In the absence of any peculiar velocities, the distortion of a spherical void in redshift space (its depth divided by its width) will be

$$e_v(z) = \frac{H(z)}{H_0} \frac{y(z)}{z} . \tag{3.1}$$

The cross-sectional area of the void, normalized to the area it would have at $z = 0$, is

$$a_v(z) = \frac{H(z)}{H_0} \frac{z}{y(z)} . \tag{3.2}$$

The volume of the void, normalized to the volume it would have at $z = 0$, is

$$V_v(z) = \frac{H(z)}{H_0} \frac{z^2}{y(z)^2} . \tag{3.3}$$

All of these functions, $e_v$, $a_v$, and $V_v$, are (in principle) measurable. The theoretical value depends on the current density parameter $\Omega_0 = (8\pi G \rho_b)/(3H_0^2)$, where $\rho_b$ is the current mean density of the mass within the universe. If the cosmological constant $\Lambda$ is not zero, the distortion will also depend on the parameter $\Omega_\Lambda = \Lambda/(3H_0^2)$. Figure 1 shows the theoretically expected values of $e_v$, $a_v$, and $V_v$ for three sets of cosmological parameters.

In the limit that $(1 + q_0)z \ll 1$,

$$e_v(z) \approx 1 + \frac{1+q_0}{2} z , \qquad (3.4a)$$

$$a_v(z) \approx 1 + 3\left[\frac{1+q_0}{2}\right] z , \qquad (3.4b)$$

and

$$V_v(z) \approx 1 + 4\left[\frac{1+q_0}{2}\right] z . \qquad (3.4c)$$

In an exponentially inflating de Sitter universe, $\Omega_\Lambda = 1$, $\Omega_0$ is negligibly small, and $q_0 = -1$. Consequently, an exponentially expanding universe shows no cosmological distortions in redshift space. However, if $\Omega_\Lambda = 0$ and $\Omega_0 \lesssim 1$, then significant cosmological distortions ($e_v \sim 2$) should occur at redshifts near unity.

To test algorithms for measuring the sizes and shapes of voids, I created a synthetic two-dimensional universe in which the galaxies were located near the cell walls of a Voronoi foam (Icke & van de Weygaert 1987). This model creates a 'bubbly' distribution for the large-scale structure, in which voids are isolated, convex structures with a characteristic length scale (Icke & van de Weygaert 1987, 1991). The number density of nuclei for the Voronoi cells in my toy universe is $\Sigma_0 = 5100(H_0/c)^2$, chosen so that the number density of cells is comparable to the number density of voids in the first CfA redshift slice (Slezak et al. 1993). The characteristic length scale for the Voronoi cells is thus $\lambda \equiv \Sigma_0^{-1/2} = 42h^{-1}$ Mpc. To make the cells more nearly uniform in size, the nuclei of the cells are anticorrelated; nuclei were laid down at random on the plane, with the constraint that no nucleus be closer than a distance $0.8\lambda$ from a previously positioned nucleus. Once the nuclei of the Voronoi cells were located, galaxies were positioned randomly, with the constraint that they lie within a distance $0.05\lambda$ of a wall dividing two cells. The number of galaxies was chosen to be 40 times the number of cells. The galaxy distribution created by this procedure is shown in the left panel of Figure 2. The mean number density of galaxies is constant out to $z = 0.3$; no observational selection effects are used. Each cell in the Voronoi foam corresponds to a void in the galaxy distribution; each wall in the Voronoi foam corresponds to a wall of galaxies between voids.

The left panel of Figure 2 shows the undistorted distribution of galaxies, as would be seen by an observer in a $q_0 = -1$ universe. The right panel shows the distorted pattern that would be seen by an observer at the origin in a $q_0 = 5$ universe. (This value of $q_0$ is much larger than we expect in our own universe; I adopt it for illustrative purposes, to make the distortions clearly visible at a modest redshift of $z \sim 0.3$). The toy universe which I have created is a particularly simple one; the voids are clearly delineated structures which are totally empty of galaxies. The method which I use to measure the size and shapes of voids must be a robust one, however – capable of coping with the more complicated distribution of voids in the real universe.

One practical algorithm for detecting voids, and measuring their size and shape, is the VOIDSEARCH algorithm of Kauffmann & Fairall (1991). When applied to a two-dimensional galaxy distribution (Kauffmann & Melott 1992), this algorithm divides the plane into a number of square cells. The largest square of empty cells is located, and the sides of the square are permitted to deform outward into adjoining empty regions. Constraints are placed on the permitted deformation to ensure that the void remains compact. The VOIDSEARCH algorithm can find the characteristic size of voids in simulations where such a characteristic scale exists; it also gives a good fit to the shape of nearly circular voids (Kauffmann & Melott 1992). Unfortunately, VOIDSEARCH does not always accurately reproduce the shape of elongated voids. The initial square of empty cells which is fitted within an elongated void can expand outward along the axes of a cartesian grid, but not in the diagonal directions. Whether the shape of an elongated void is accurately measured thus depends on whether or not it happens to be aligned with the superimposed grid of cells.

Since I want to measure the elongation of cells as accurately as possible, I have adopted an algorithm which is different from the VOIDSEARCH algorithm. Instead of inscribing squares within voids, I inscribe ellipses within voids, with the constraint that one of the principal axes of the ellipse must lie along the line of sight from the origin. The axis ratio of the ellipse then provides an estimate of the elongation $e_v$ of the void.

Let me now outline how my void-finding algorithm works. I start by superimposing a grid of points upon the galaxy distribution in redshift space. To provide adequate resolution of voids, the distance between adjacent grid points must be small compared to the characteristic size of voids. In the toy universe examined in this paper, the characteristic void size is $\delta z \sim 0.014$ for nearby voids. The superimposed grid which I used has $\delta z \sim 0.0008$. For each grid point, I find the largest ellipse which is centered on the point and which contains no galaxies; one of the principal axes of the ellipse is constrained to lie along the line of sight from the origin to the grid point. I allow the axis ratios of the ellipse to vary from $10^{-1/2}$ to $10^{1/2}$, and determine the axis ratio which maximizes the area of the vacant ellipse. The constraint that the vacant ellipse be aligned with the line of sight is imposed because I am interested solely in the radial distortions in redshift space. A more general fitting procedure, if the radial distortions were not the main point in question, would allow the axes of the ellipse to be aligned arbitrarily.

For each grid point, I thus find the area and the axis ratio of the largest vacant ellipse centered on that point. I rank all of the vacant ellipses which I have found according to their area. The largest ellipse is designated the largest void in the sample. I test the second largest ellipse to see whether it overlaps with the largest ellipse. If it overlaps, it is discarded; if it doesn't overlap, it is designated the second largest void in the sample. This procedure is repeated, in turn, with all of the ellipses in the list, in order of decreasing area. If the tested ellipse overlaps with any of the previously designated voids, it is discarded. If it doesn't overlap, it is added to the list of voids.

The end result of this procedure is a list of non-overlapping ellipses, empty of galaxies, whose principal axes are aligned with the line of sight from the origin, and whose areas are maximized in a well-defined manner. These ellipses are labeled "voids". (A similar procedure for a three-dimensional distribution of galaxies would define voids as non-overlapping ellipsoids whose axes of symmetry lie along the line of sight from the origin.) Figure 3 displays the largest elliptical voids found in the toy universe; the left panel assumes $q_0 = -1$ and the right panel assumes $q_0 = 5$. Only voids whose centers are at $z < 0.26$ are shown, ignoring the region where edge effects are important due to the abrupt cutoff in galaxies at $z = 0.3$.

The ellipse-fitting algorithm is a simple-minded one, and takes no notice of the fact that there is a characteristic void size in the toy universe. Once it fits an elliptical void to each of the Voronoi cells (compare Figure 2 and Figure 3), it goes on and fits smaller and smaller ellipses into the available vacant corners. Figure 3 shows only those elliptical voids whose dimensionless area is greater than $8 \times 10^{-5}$; the smaller ellipses, those which are tucked into the corners of the Voronoi cells, are not displayed. A cursory glance at Figure 3 demonstrates that the area of voids and the elongation of voids both increase with redshift in a $q_0 = 5$ universe. However, if the value of $q_0$ is in the more plausible range of $-1 \lesssim q_0 \lesssim 1$, the distortions will be more subtle. It will be useful to know, how deep a redshift survey must go in order to distinguish between a spatially flat universe in which matter dominates ($q_0 = 1/2$) and a flat universe dominated by a cosmological constant ($q_0 = -1$).

For my two-dimensional toy universe, Figure 4 plots the area of the 1800 largest elliptical voids with centers at $z < 0.26$. The theoretical value of $\Sigma_0 a_v(z)$, as given by equation (3.4b), is shown as the dashed line in each panel. Most of the elliptical voids fall below the dashed line. This is to be expected: it merely says that the ellipses are smaller in

– 9 –

area than the polygonal Voronoi cells in which they are inscribed. The 'true' elliptical voids, those with areas greater than $8 \times 10^{-5} \approx 0.4/\Sigma_0$, do a fair job of tracking the increasing area of the voids with redshift. The solid lines in Figure 4 shows the best least-squares fit of the form

$$\text{Area} = \alpha_0 + \alpha_1 z \; , \tag{3.5}$$

using only those voids with areas larger than $0.4/\Sigma_0$. In the $q_0 = -1$ universe, the least-square estimators, with standard errors, are $\alpha_0 = (1.38 \pm 0.02) \times 10^{-4}$ and $\alpha_1 = (-0.28 \pm 0.12) \times 10^{-4}$, with a coefficient of correlation $r = -0.068$. In the $q_0 = 1/2$ universe, $\alpha_0 = (1.33 \pm 0.04) \times 10^{-4}$ and $\alpha_1 = (3.59 \pm 0.21) \times 10^{-4}$, with $r = 0.54$.

From the slope and intercept of the best least-squares fit, an estimate

$$q_a = \frac{2}{3} \frac{\alpha_1}{\alpha_0} - 1 \tag{3.6}$$

can be made for the deceleration parameter $q_0$. If the errors in $\alpha_0$ and $\alpha_1$ are assumed to be Gaussian, then the estimate, with standard error, is $q_a = -1.13 \pm 0.06$ in the $q_0 = -1$ universe and $q_a = 0.80 \pm 0.09$ in the $q_0 = 1/2$ universe. Thus, in the $q_0 = 1/2$ universe, the estimate of $q_0$ is too large at about the $3\sigma$ confidence level. If the wall between two adjoining voids is not detected for any reason (for instance, if the galaxies which formed within the wall are all too faint to be included within a particular redshift survey), then the void-fitting algorithm will treat the two adjacent voids as a single large void, and find one oversized ellipse within the area.

Failure to detect an authentic void wall will result in the identification of an artificial void which is larger than the characteristic void size at that redshift (assuming that there is such a characteristic size). Thus, using the size of voids to estimate $q_0$ is no longer a valid procedure at redshifts where the number of detected galaxies is too small to accurately define the void walls. To illustrate this point, I created a two-dimensional Voronoi universe which has only 5 galaxies per Voronoi cell (let's call it the 'sparsely sampled' universe, since it contains so few galaxies around each void). The sparsely sampled universe was created by starting with the universe shown in Figure 2 (which has 40 galaxies per cell) and randomly removing seven-eighths of the galaxies. In a Voronoi universe with only 5 galaxies for every cell, the mean number of galaxies per cell wall is 5/3. Poisson statistics will therefore result in many void walls which contain no galaxies at all.

Applying the void-fitting algorithm to the sparsely-sampled universe yields the results shown in Figure 5. Note that the characteristic void size is no longer easily detected. There is no longer a sharp distinction between the 'true' elliptical voids, which nearly fill a Voronoi cell, and the smaller ellipses that are tucked into the cell corners. The lesson to be learned from comparing Figures 4 and 5 is that the characteristic void area is easier to detect, and hence $q_0$ is easier to estimate, when the large-scale structure is well-defined, with several galaxies within each void wall, and no stray galaxies within voids.

An estimate of the deceleration parameter $q_0$ can also be made by measuring the axis ratio $e_v$ of each fitted ellipse, instead of its area. Figure 6 is a plot of the natural logarithm of $e_v$ as a function of the redshift $z$; only the elliptical voids with area greater than $0.4/\Sigma_0$ are included. The dashed line in each panel is the expected relation

$$\ln e_v = \frac{1+q_0}{2} z \qquad (3.8)$$

which is expected for circular voids when $(1+q_0)z \ll 1$. Since the voids are not intrinsically spherical, there is a significant scatter in $e_v$ for a given redshift. In the $q_0 = -1$ universe, the standard deviation in $\ln e_v$ is approximately 0.25.

Despite the scatter in the intrinsic axis ratios, the trend for increasing elongation at greater redshifts can be picked out of the data. For both the $q_0 = -1$ universe and the $q_0 = 1/2$ universe, I made a least-squares fit to the data, using the functional form

$$\ln e_v = \beta_0 + \beta_1 z \qquad (3.9)$$

For the $q_0 = -1$ universe, the best fit (with standard errors) had $\beta_0 = 0.013 \pm 0.024$ and $\beta_1 = -0.034 \pm 0.128$, with correlation coefficient $r = -0.008$. For the $q_0 = 1/2$ universe, the best fit gave $\beta_0 = 0.017 \pm 0.027$ and $\beta_1 = 0.692 \pm 0.151$, with $r = 0.54$. The best fits to equation (3.9) are shown as the solid lines in Figure 6, nearly indistinguishable on this scale from the theoretically expected dashed lines.

An estimate of the cosmological deceleration parameter, comparing equations (3.8) and (3.9), is

$$q_e = 2\beta_1 - 1 \; . \qquad (3.10)$$

In the $q_0 = -1$ toy universe, the estimated value of $q_0$, derived from the axis ratio of the fitted elliptical voids, is $q_e = -1.07 \pm 0.26$. For the $q_0 = 1/2$ universe, $q_e = 0.38 \pm 0.30$. Thus, measurements of the axis ratio are capable (in a toy universe with no peculiar velocities) of distinguishing between a flat universe which is matter-dominated and one which is dominated by a cosmological constant. However, the scatter in the intrinsic axis ratios of voids leads to a fairly large uncertainty in the estimate of $q_0$ derived by this method.

How deep must a redshift survey go in order to yield an accurate estimate of $q_0$? A survey with a depth of $z \lesssim 0.1$ will contain only a small number of voids and will display only a small cosmological distortion. A deeper survey is necessary to distinguish between a universe with $q_0 = 1/2$ and one with $q_0 = -1$. To give some indication of how deep the survey must go, I computed the estimated value of $q_0$ in the toy universe as a function of 'survey depth' – that is, I estimated $q_0$ using only those voids with redshifts less than some limiting value $z_d$. The two-dimensional toy universe corresponds to a thin 360° slice in the real universe.

Figure 7 is a plot of the estimated value of $q_0$ as a function of the depth $z_d$ in redshift space. On the left, the estimates $q_a$, drawn from the areas of the fitted elliptical voids, are shown. On the right, the estimates $q_e$, drawn from the axis ratios, are shown. In addition to the best estimate, shown as the solid line, each panel shows the standard errors around the best estimate (the dotted lines above and below the solid line). The errors decrease in size with increasing redshift. At a depth of $z_d = 0.05$, which contains 40 voids in the $q_0 = -1$ universe and 32 voids in the $q_0 = 1/2$ universe, there are simply too few voids to accurately determine $q_0$.

## 4. Discussion

In an idealized universe, where voids are clearly delineated by galaxies along their walls and where the galaxies have no peculiar velocities, it is possible to estimate the value of $q_0$ from the distortion of voids in redshift space. Even in an idealized universe, a redshift slice would have to have a depth $z \gtrsim 0.1$ in order to distinguish between a $q_0 = 1/2$ and a $q_0 = -1$ universe. Distinguishing between $q_0 = 1/2$ and $q_0 = 0$ would require an even deeper survey.

Alcock and Paczyński (1979) proposed using the ratio of depth in redshift space ($\delta z$) to width in redshift space ($z \delta \theta$) as a means of measuring $q_0$. One can as easily use areas in redshift space (depth times width) or, in a three-dimensional survey, volumes in redshift space (depth times the square of the width). If voids were all intrinsically spherical, but with different intrinsic volumes, then the axis ratio in redshift space, $e_v$, would give the best estimate of $q_0$. Conversely, if the spheres were intrinsically ellipsoidal, but with identical intrinsic volumes, then the volume in redshift space, $V_v$, would give the best estimate of $q_0$. Since real voids have a scatter both in size and in shape, all estimators for the value of $q_0$ – $e_v$, $a_v$, and $V_v$ – will contain uncertainties.

A great advantage of using the methods outlined in this paper is that, unlike the 'standard candle' methods, they do not strongly depend on the luminosity evolution of galaxies. As long as the galaxies are bright enough to be included in the redshift survey, their precise luminosities and colors are irrelevant. The survey must simply contain enough galaxies to accurately delineate the void walls. In addition, for the purposes of this paper, the origin of the voids is largely irrelevant. Whether they form by gravitational forces alone, or were triggered by explosions, they have stamped a convenient pattern upon the universe, whose distortions can be measured.

The origin of the voids is only relevant in the extent to which peculiar velocities are imparted to the galaxies. For instance, numerical simulations of an $\Omega_0 = 1$ universe show that large voids have excess expansion velocity of $F \approx 0.15 - 0.20$ in a CDM universe, but only $F \approx 0.05 - 0.08$ in an HDM universe (van de Weygaert & van Kampen 1993). If voids expand more rapidly than the Hubble flow, they will possess an additional elongation in redshift space, beyond the cosmological elongation. Studies of voids at $z \lesssim 0.05$ (Bothun et al. 1992, Slezak et al. 1993) indicate that the distortion of voids due to peculiar velocities is small. An additional effect of peculiar velocities is to convert virialized clusters into 'fingers of god' in redshift space. These fingers, by poking into the near side and far side of voids, decrease the volume of the largest ellipsoid which can be fit into a void. If the ratio of the average finger length to the average void radius is roughly constant with time, then the effect of virialized clusters can be more easily compensated for. In any case, the peculiar velocities associated with clusters will prove the greatest stumbling block to measuring the distortion of real voids in redshift space, and thereby measuring the value of $q_0$.

I thank Tod Lauer for introducing me to the work of Alcock & Paczyński, and for encouraging me to undertake this study. Cheongho Han provided invaluable calculations and discussions.

Fig. 1.— The top panel shows the line-of-sight distortion $e_v$ in redshift space; the short-dashed line indicates a universe with $\Omega_0 = 1$ and $\Omega_\Lambda = 0$, the solid line indicates a universe with $\Omega_0 = 0.1$ and $\Omega_\Lambda = 0$, and the long-dashed line indicates a universe with $\Omega_0 = 0.1$ and $\Omega_\Lambda = 0.9$. The middle panel shows the cross-sectional void area in redshift space, and the bottom panel shows the void volume $V_v$ in redshift space; the meaning of line types is the same as in the upper panel.

Fig. 2.— The distribution of galaxies in a toy universe in which the voids are the cells of a Voronoi foam. The left panel shows the redshift map for an observer located at the origin in a universe with $q_0 = -1$. The right panel shows a similar map for a universe with $q_0 = 5$.

Fig. 3.— The elliptical voids which are fitted to the toy universes shown in Figure 2. Voids whose centers are at $z < 0.26$ and whose dimensionless area is $a > 0.4/\Sigma_0$ are shown. In the $q_0 = -1$ universe (left panel) there are 1059 such voids; in the $q_0 = 5$ universe (right panel) there are 351.

Fig. 4.— The dimensionless area of the 1800 largest elliptical voids whose centers are at a redshift $z < 0.26$. The top panel assumes $q_0 = 1/2$; the bottom panel assumes $q_0 = -1$. In each panel, the dashed lines shows the theoretically expected value for $\Sigma_0 a_v(z)$; the solid line is the best least-squares fit for those voids with area greater than $0.4/\Sigma_0$.

Fig. 5.— The same results as the previous Figure, but for a universe with only 5 galaxies per void, instead of 40 galaxies per void. Because of the much larger scatter in areas, a least-squares fit was not made.

Fig. 6.— The natural logarithm of the void elongation; only voids with area greater than $0.4/\Sigma_0$ are included. The top panel assumes $q_0 = 1/2$ and the bottom panel, $q_0 = -1$. The dashed line shows the theoretically expected value; the solid line shows the best least-squares fit.

Fig. 7.— The estimated value of $q_0$ as a function of survey depth in redshift space. The left-hand panels show $q_a$, the value estimated from the void areas; the right-hand panels show $q_e$, the value estimated from the axis ratios of the voids. The heavy solid line is the estimated value; the dotted lines are the estimated value plus or minus the standard error. The horizontal line in each panel shows the true value of $q_0$.